# Oxygen-induced *in-situ* manipulation of the interlayer coupling and exciton recombination in Bi$_2$Se$_3$/MoS$_2$ 2D heterostructures

Zachariah Hennighausen,[1] Christopher Lane[1], Abdelkrim Benabbas[3], Kevin Mendez[1], Monika Eggenberger[1], Paul M. Champion[2], Jeremy T. Robinson[3], Arun Bansil[1], and Swastik Kar[1,*]

[1] Department of Physics, Northeastern University, Boston, MA 02115
[2] Physics Department and Center for Interdisciplinary Research on Complex Systems, Northeastern University, Boston, MA 02115
[3] Naval Research Laboratory, Washington, DC 20375

*Author for correspondence, Email: s.kar@northeastern.edu

**2D heterostructures are more than a sum of the parent 2D materials, but are also a product of the interlayer coupling, which can induce new properties. In this paper we present a method to tune the interlayer coupling in Bi$_2$Se$_3$/MoS$_2$ 2D heterostructures by regulating the oxygen presence in the atmosphere, while applying laser or thermal energy. Our data suggests the interlayer coupling is tuned through the diffusive intercalation and de-intercalation of oxygen molecules. When one layer of Bi$_2$Se$_3$ is grown on monolayer MoS$_2$, an influential interlayer coupling is formed that quenches the signature photoluminescence (PL) peaks. However, thermally annealing in the presence of oxygen disrupts the interlayer coupling, facilitating the emergence of the MoS$_2$ PL peak. DFT calculations predict intercalated oxygen increases the interlayer separation ~17%, disrupting the interlayer coupling and inducing the layers to behave more electronically independent. The interlayer coupling can then be restored by thermally annealing in N$_2$ or Ar, where the peaks will re-quench. Hence, this is an interesting oxygen-induced switching between "non-radiative" and "radiative" exciton recombination. This switching can also be accomplished locally, controllably, and reversibly using a low-power focused laser, while changing the environment from pure N$_2$ to air. This allows for the interlayer coupling to be precisely manipulated with submicron spatial resolution, facilitating site-programmable 2D light-emitting pixels whose emission intensity could be precisely varied by a factor exceeding 200×. Our results show that these atomically-thin 2D heterostructures may be excellent candidates for oxygen sensing.**

Research in 2D heterostructures continues to gain enormous interest for their potential to advance both fundamental and application-oriented research. They have contributed to fields as diverse as transistors,[1]–[3] optoelectronics,[4]–[6] information storage,[7]–[9] plasmonics,[10]–[12] photocatalysis,[13]–[15] capacitors,[16], [17] biosensors,[18] spintronics,[19], [20] high-density lithium storage,[21], [22] and superconductivity.[23]–[25] The far-reaching success of 2D heterostructures is in part due to the large spectrum of properties they have demonstrated. A 2D heterostructure's properties are more than a sum of the parent 2D materials, but are also a product of the interlayer interaction, which can be manipulated to engineer new capabilities. It has been shown that the twist angle,[26] interlayer spacing,[25] thermal annealing,[27]–[29] and intercalation of molecules[30]–[32] all influence the interlayer coupling. There are no tools to directly probe the interlayer coupling strength; however, its impact can be inferred by the extent it influences the properties. In this paper we demonstrate that the interlayer coupling in $Bi_2Se_3$/$MoS_2$ 2D heterostructures can be tuned by regulating the oxygen presence in the atmosphere, while applying controlled laser or thermal energy doses. The coupling strength is inferred using changes in the photoluminescence (PL) intensity, where lower PL corresponds to higher coupling. Our data suggests the interlayer coupling is modulated by diffusively intercalating and de-intercalating oxygen molecules.

The interlayer coupling in 2D materials is considered a promising parameter for designing materials with tailored properties; however, despite the significant interest, the interlayer coupling is not well understood, which is in part due to the lack of experimental techniques with the ability to precisely manipulate it. Some previous experimental work has focused on using global thermal annealing to manipulate the coupling; however, this method has not demonstrated precision, and often requires several hours to complete.[27]–[29] Ion irradiation,[33] in-plane strain,[34] and lateral pressure[25], [35] have all been shown to increase the interlayer coupling by decreasing the interlayer separation; however, ion irradiation has also been shown to damage 2D materials and induce defects[36]–[38], and applying strain or pressure alters the 2D material's lattice parameters and properties,[39]–[41] thereby introducing uncertainty. Work to manipulate the interlayer coupling via electric fields has only been theoretical, and follow-up experimental work is needed to confirm it.[42]–[45] In this paper we demonstrate a facile method to tune the interlayer coupling *in-situ* with high-spatial resolution, all using ambient conditions and tools commonly found in a 2D research laboratory.

Monolayer $MoS_2$ is known, among other things, for having a bright PL due to the formation of tightly-bound excitons.[46] However, when only one layer of $Bi_2Se_3$ is grown on the monolayer $MoS_2$ using vapor-phase deposition, the bright PL is >99% quenched because the interlayer coupling induces a non-radiative exciton recombination pathway.[7] In this paper we demonstrate that the PL can be controllably increased and decreased at small increments of only a few percent, suggesting that the coupling is being diminished or strengthened, respectively, at the same small increments. This is done by applying energy to the material in either an oxygen-present or oxygen-absent environment, which we believe facilitates the diffusive intercalation or de-intercalation of oxygen molecules. Our results suggest that the diffusive intercalation of oxygen molecules induces the monolayer $MoS_2$ to behave more electronically independent, thereby restoring the radiative recombination pathway and facilitating the emergence of the signature PL peak.

The intercalation of molecules between layers has been shown to disrupt their coupling, leading to change in the properties.[21], [30]–[32], [47]–[53] For example, it has been shown that when several layer $MoS_2$ is intercalated with lithium or quaternary ammonium molecules, the signature monolayer photoluminescence emerges, suggesting the intercalated molecules induce the $MoS_2$ layers to behave electronically independent, as if they were monolayer.[30], [47], [54] It has also been shown that oxygen can easily intercalate between 2D crystals and their substrates, decoupling the two materials and inducing them to behave more "freestanding" (*i.e.* electronically independent).[31], [32], [48]–[52]

A natural question arises as to why oxygen, vice other molecules in the atmosphere, is manipulating the interlayer coupling in $Bi_2Se_3/MoS_2$ 2D heterostructure. $Bi_2Se_3$ is well-known for its oxygen-affinity. It has been shown that initial absorption of oxygen p-dopes the crystal,[55]– [57] gradually destroying its exotic topological properties and eventually forming wide-band bismuth and selenium oxides.[56] Indeed, the deleterious oxidation of $Bi_2Se_3$ has so far restricted most practical applications using $Bi_2Se_3$.

In this work, we obtain highly controllable, reversible, and site-selectable switching between direct (*i.e.* radiative and photoluminescent) and indirect (*i.e.* non-radiative) exciton-recombination pathways in $Bi_2Se_3$/$MoS_2$ 2D heterostructures (1-3 layers of $Bi_2Se_3$ grown on monolayer $MoS_2$). Our approach allows permanent, *in-situ*, electrode-less, and use-specific programming of the interlayer coupling and PL intensity. Exciton recombination dynamics could be switched (between radiative and non-radiative) by annealing the heterostructures in oxygen-present (*e.g.* air) *vs.* inert atmospheres (*i.e.* Ar or $N_2$). Alternately, the switching could be localized in a highly controllable manner at ambient temperatures using a focused laser (as before, in air or under $N_2$), which allowed site-selective reversible manipulation of different regions of the same 2D heterostructures. Additionally, our results suggest that these 2D heterostructures might have applications as standard temperature-pressure high-density oxygen storage devices, potentially storing 69 kg/m$^3$ (a factor of 52 times the density of $O_2$ gas at 1 atm).

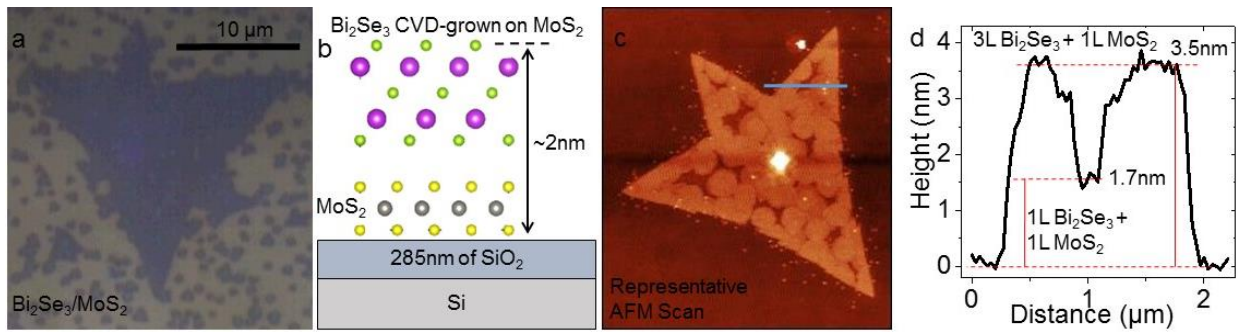

**Figure 1. As-grown $Bi_2Se_3$/$MoS_2$ vertical 2D heterostructures** (a) Optical image of a $Bi_2Se_3$/$MoS_2$ vertical 2D heterostructure, where 1 layer of $Bi_2Se_3$ was uniformly grown on a monolayer $MoS_2$ crystal using vapor-phase deposition. (b) Side-view diagram of the heterostructure. (c) Representative atomic force microscope (AFM) scan of a $Bi_2Se_3$/$MoS_2$ 2D heterostructure, demonstrating monolayer $Bi_2Se_3$ (with trilayer islands) grew uniformly across the entire $MoS_2$ crystal.

Figure 1a shows an optical image of $Bi_2Se_3$/$MoS_2$ 2D heterostructure (1 layer $MoS_2$ + 1 layer $Bi_2Se_3$)

on SiO$_2$, where the monolayer MoS$_2$ was grown using vapor-phase chalcogenization (VPC)[58] and the Bi$_2$Se$_3$ was grown on top using vapor-phase deposition. Figure 1b is a side-view diagram of a typical Bi$_2$Se$_3$/MoS$_2$ 2D heterostructure. Despite the huge lattice mismatch (*e.g.* 2.74Å to 3.57Å, see SI.1), uniform layers of Bi$_2$Se$_3$ grow with high regularity on top of the MoS$_2$ crystal, suggesting strong van der Waals epitaxy-mediated growth between the two component layers.[59] SI.2 shows the transfer characteristics of back-gated monolayer MoS$_2$ and Bi$_2$Se$_3$/MoS$_2$ FET devices. The right-shift of the threshold gate voltage in the 2D heterostructure indicates a relative downshift of the effective Fermi level due to reduction of excess n-type carriers (by about ~9.50x10$^{12}$ cm$^{-2}$). The device response looks neither like monolayer MoS$_2$, nor few layer Bi$_2$Se$_3$ (a semi-metal),[60] suggesting there is sufficient interlayer coupling to modify the electronic structure of both materials.

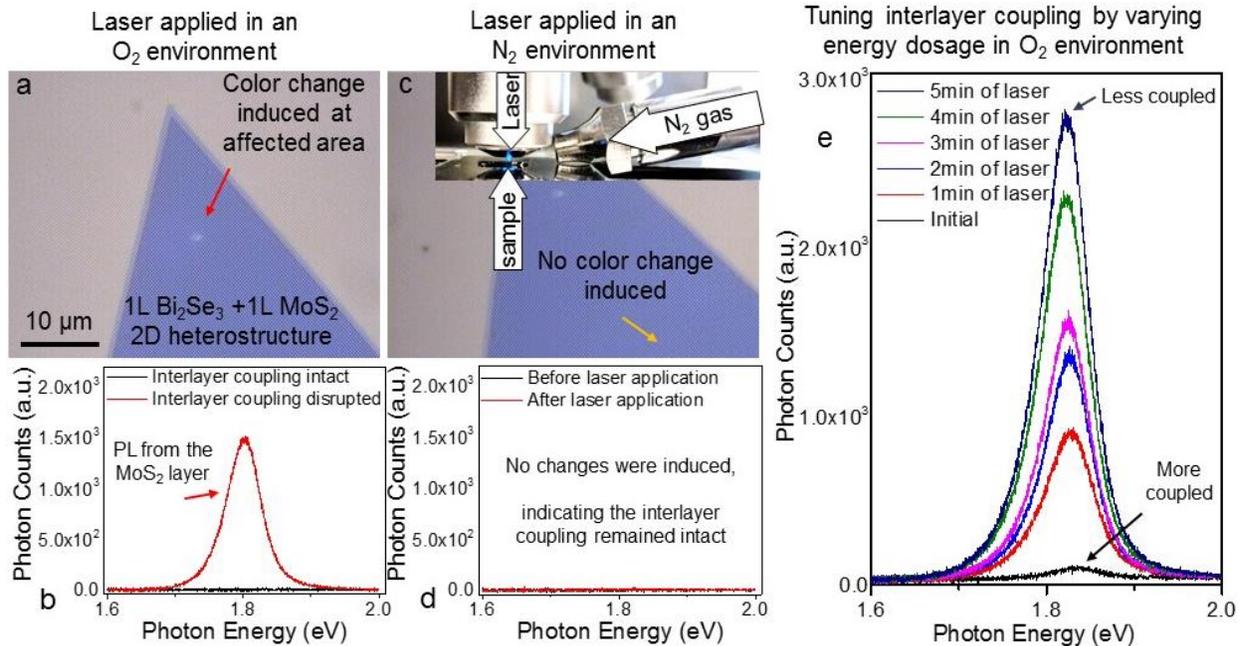

**Figure 2. Manipulating the interlayer coupling using oxygen and energy** (a) Optical image of a $Bi_2Se_3$/$MoS_2$ vertical 2D heterostructure where energy was applied locally using a focused laser, while the sample was in an oxygen-present environment (*i.e.* air). The optical properties of the affected area were altered going from purple to white, allowing affected locations to be easily identified. (b) PL spectra from the same spot before and after energy was applied, demonstrating how the interlayer coupling can be manipulated. Initially, the PL spectra was flat; however, after applying a focused laser for 8 minutes at 168μW, a PL spectra corresponding to monolayer $MoS_2$ appeared, suggesting the interlayer coupling was disrupted, allowing the $MoS_2$ layer became more electronically independent. Electronically independent monolayer $MoS_2$ has a signature PL peak due to its tightly bound excitons. (c) The laser was now applied to a different location (orange arrow) on the same sample at the same power (168μW for 12 minutes), but the environment was changed from oxygen to nitrogen. Interestingly, the color change seen in (a) was not observed. The inset shows the setup where $N_2$ gas was flown across the sample to displace the oxygen. (d) PL spectra from before and after energy was applied show no perceivable appearance of PL, suggesting the interlayer coupling was not affected. Thermal annealing experiments in SI.4 demonstrate applying heat in an oxygen environment disrupts the interlayer interaction, and that the other components in air (*e.g.* $N_2$, $H_2O$, $CO_2$) do not appear to affect the interlayer coupling. (e) PL spectra of a $Bi_2Se_3$/$MoS_2$ 2D heterostructure before energy was applied, as well as after several consecutive doses, demonstrating how the PL intensity is dependent on the total dose.

Figure 2a is an optical image of a $Bi_2Se_3$/$MoS_2$ vertical 2D heterostructure, where energy was applied locally using a focused laser (168μW for 8 minutes), while the sample was in an oxygen-present environment (*i.e.* air). The red arrow identifies the affected location, which underwent a change in color from purple to white. Figure 2b shows its PL spectra before and after the energy dose was applied. Characteristic $B_2Se_3$/$MoS_2$ 2D heterostructures do not have a PL, or only a very weak PL, because the interlayer coupling creates a non-radiative recombination path for the tightly-bound excitons that exist in the monolayer $MoS_2$.[7] However, if energy is applied to the heterostructure in the form of either thermal annealing or a laser, while the heterostructure is in an $O_2$-present environment, a PL spectrum emerges that corresponds to that of monolayer $MoS_2$, suggesting the interlayer coupling was disrupted and the signature $MoS_2$ excitons are recombining along radiative pathways. Atomic force microscope (AFM) measurements demonstrate that the $Bi_2Se_3$ remains on the $MoS_2$ after a color change has been induced and the PL remerges, verifying that the change is not due simply to the removal of $Bi_2Se_3$ (SI.3). A different spot on the same sample (Figure 5c, yellow arrow) was exposed to the same power (168μW for 12 minutes), while being continuously purged with $N_2$ gas (inset Figure 5c), thereby removing oxygen from the vicinity of the exposed spot. We find that even with an increased laser dose (12 min. *vs.* 8 min. exposure), there was neither a perceivable color change, nor emergence of the PL peak (Figure 5d), suggesting the interlayer coupling was not affected. Next, the atmosphere was switched back to air without changing the laser power or moving the sample - thereby re-introducing oxygen without changing the specific location being probed - resulting in a color change and a strong PL peak recovery (SI.4b), confirming that oxygen plays a critical role in the observed changes. Detailed investigations in different gas environments in SI.4 establish that oxygen ($O_2$), and not nitrogen, $H_2O$ vapor, or carbon dioxide, is required to induce the changes and manipulate the interlayer coupling. Figure 2e demonstrates how the PL intensity is dependent on the energy dose applied, suggesting that the interlayer coupling can be manipulated in small increments. Later in the paper we demonstrate that the interlayer coupling can be tuned with high precision by controlling both the energy applied and the partial pressure of oxygen in the surrounding atmosphere.

Photo-excited e-h pairs in mono-layered transition metal dichalcogenides (TMDs) form tightly-bound neutral and charged excitons.[46] In direct-gap monolayer TMDs, they recombine radiatively, producing well-known PL spectra. In few-layered and thicker TMDs the quasiparticle band gaps are indirect, hence the K→Γ indirect (non-radiative) recombination pathway becomes more favorable, resulting in progressively suppressed PL.[61] Indirect recombination in certain heterostructures can similarly be non-radiative, if the excitons formed near a $\vec{k}$-vector in the reciprocal lattice of one layer finds the most favorable recombination pathway via a lower-energy-state that is located at a different $\vec{k}$-point in the reciprocal lattice of the second layer. In all the as-grown Bi$_2$Se$_3$/MoS$_2$ 2D heterostructures, the PL spectra were strongly quenched, suggesting the interlayer coupling was inducing non-radiative recombination pathways for the excitons. Upon application of energy (either heat or laser) in an oxygen-present environment, all heterostructures recovered their radiative recombination pathways, suggesting the interlayer coupling was disrupted, thereby inducing the MoS$_2$ layer to behave more electronically independent. Next we show that the PL can be quenched again by applying energy (via either thermal annealing or laser exposure) in an O$_2$-free environment (SI.4), demonstrating that the effects could be reversed and the interlayer coupling restored.

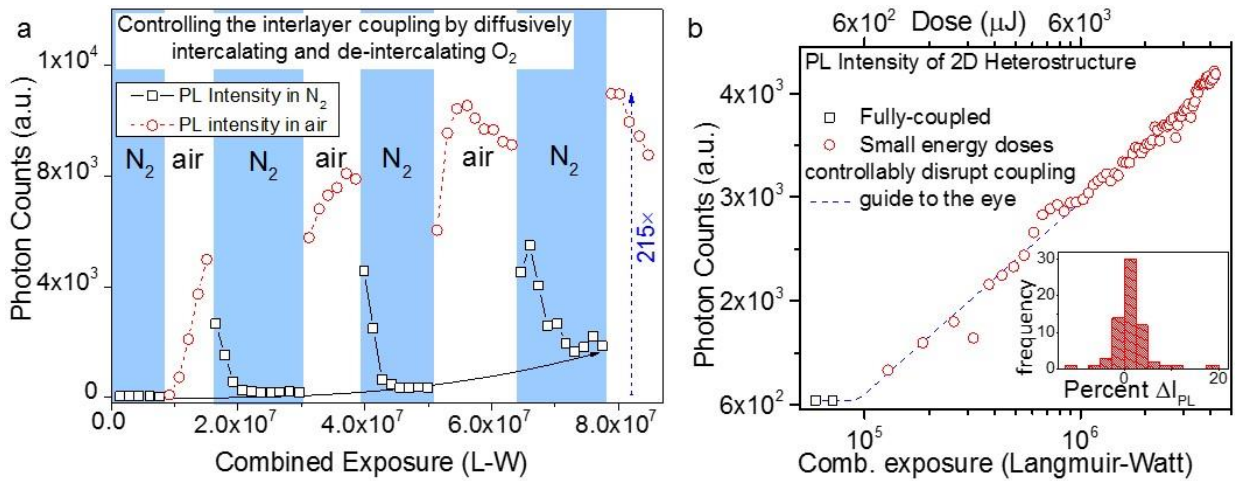

**Figure 3. Tuning the interlayer coupling and photoluminescence intensity** (a) Variation of PL intensity under alternating air and nitrogen environments while a focused laser (*i.e.* energy) is applied, demonstrating that a nitrogen environment is able to restore the interlayer coupling. The solid black arrow shows how the baseline PL reading monotonically grows after repeated environment cycling, and the PL's rate-of-change varies, suggesting the 2D heterostructure may not return to its initial (as-grown) state. The dashed blue arrow shows the overall PL intensity growth factor, up to 215×, achievable by this cycling approach. (b) Correlation between laser (*i.e.* energy) exposure (in air) and the resulting PL intensity of a Bi$_2$Se$_3$/MoS$_2$ 2D heterostructure using a calibrated recipe (50 μW, 6 s doses followed by 1 μW, 60 s for collecting data), demonstrating a controlled disruption of interlayer coupling. Under this dose recipe, the radiative recombination was found to grow approximately logarithmically, and a vast majority of exposures result in a <5% change in intensity (see inset). See also SI.5 for other possible behaviors when the recipe is changed. Inset: histogram of the change steps under each exposure. Both (a) and (b) taken together demonstrate an unprecedented degree of controlled manipulation of the interlayer coupling and PL achievable. Later we demonstrate this is possible in a site-selectable manner as well.

Remarkably, not only are we able to disrupt the interlayer coupling and increase the PL intensity using, but we can also restore the interlayer coupling and decrease the PL intensity, which we believe is due to the intercalation and de-intercalation of oxygen. Figure 3a shows the change in PL intensity of a $Bi_2Se_3$/$MoS_2$ 2D heterostructure that is oxygenated and deoxygenated several cycles by switching the ambient atmosphere between air and nitrogen. This controllable switching of radiative and non-radiative exciton recombination pathways, demonstrates the ability to both disrupt and restore the interlayer coupling. This process can be cycled several times, underscoring the fact that at the initial stage, under identical laser power, the oxygenation is reversed by the mere removal of the $O_2$ partial pressure in the ambient. This indicates that at least initially, the oxygenation process is diffusive and does not form chemical bonds. After a few cycles, the maximum PL intensity grows by as much as 215× and stops quenching fully, suggesting that other more permanent changes occur at higher energy dosage (discussed later in the paper).

Figure 3b shows the variation of PL intensity measured after repeated doses (t=6s) of combined exposure to air (at ambient pressure) and laser power (at 50 µW). At this dose-value, the PL was found to grow approximately logarithmically with combined exposure (along with incident energy), suggesting the interlayer coupling can be tuned with high precision. The logarithmic shape is in agreement with Fick's law of diffusion, which states the diffusion flux will decrease as the system reaches equilibrium, and whose general solution has an exponential form. Additionally, Fick's law has been used to describe the intercalation process, including the intercalation of lithium into vertically stacked bilayer graphene.[22] The inset shows that the intensity-change ($\Delta I_{PL}$) can be as low as~5%, suggesting possible applications as low-cost atomically-thin laser calorimeters or photon-counters. SI.5 demonstrates other possible behaviors when the recipe is changed.

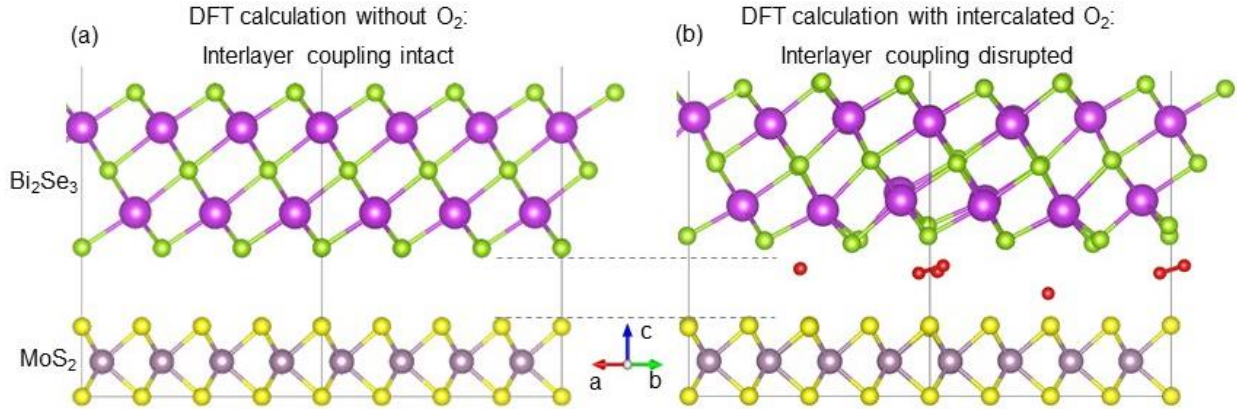

**Figure 4. DFT calculations predict intercalated O₂ diminishes interlayer coupling** (a) DFT calculations of a rotationally aligned (*i.e.* twist angle is 0°) $Bi_2Se_3$/$MoS_2$ superlattice predict significant charge redistribution into the interlayer region, and an influential interlayer coupling. (b) However, when O₂ molecules are placed in the interface between the layers, DFT calculations predict the average interlayer separation increases from 3.57Å to 4.18Å (17% increase), diminishing the interlayer coupling. These results are in agreement with previous studies where intercalated O₂ between a 2D material and the substrate induces the 2D material to behave electronically independent (*i.e.* "freestanding"). Further, intercalation and de-intercalation are diffusion-dependent processes, offering an explanation for why the PL intensity rises and falls so quickly as the environment changes (Figure 3). DFT calculations predict an interlayer separation ~4.7 times larger than the diameter of an oxygen atom, suggesting sufficient space exists for O₂ molecules to reside.

The rapid changes observed in Figure 3 by simply switching the environment between air (*i.e.* $O_2$-present) and nitrogen (*i.e.* $O_2$-absent) suggest that oxygen is diffusing in and out of the 2D heterostructure. As seen in Figure 4, the $O_2$ molecule is relatively small compared to the interlayer spacing. Further, it has been shown that $O_2$ molecules are able to permeate into $Bi_2Se_3$, demonstrating they are able to fit between the atoms.[56], [57] Density functional theory (DFT) calculations predict that the interlayer coupling in a $Bi_2Se_3$/$MoS_2$ 2D heterostructure induces significant charge redistribution in the interlayer region, hybridizing nearest neighbor atoms to form bonds, and that it is influential in modifying the bandstructure.[7] Interestingly, when $O_2$ molecules are placed in the interlayer region, DFT calculations predict that the materials will begin to separate, diminishing the interlayer coupling, and inducing the layers to behave more electronically independent. It can be seen in Figure 4 that the interlayer spacing increases on average by 17% when five $O_2$ molecules are intercalated. Together these results hint at the interesting possibility that these atomically-thin layers may be excellent candidates for oxygen storage devices, potentially storing 69 kg/m³ (a factor of 52 times the density of $O_2$ gas at 1 atm).

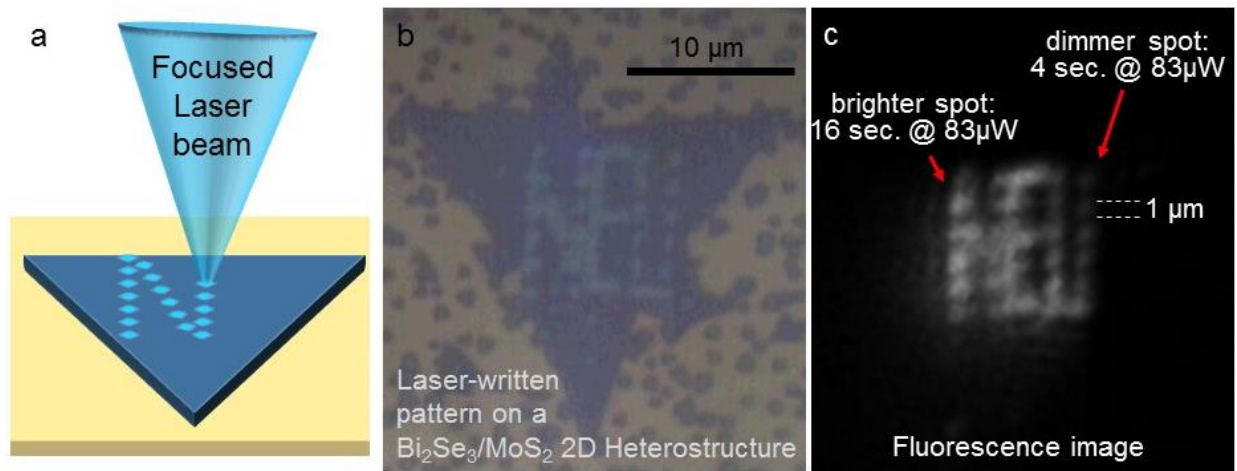

**Figure 5. Site-selective manipulation of interlayer coupling and photoluminescence** (a) Schematic of a method to write patterns with site-selected interlayer coupling strength on $Bi_2Se_3/MoS_2$ 2D heterostructures using a focused laser. (b) Optical image of a laser-written pattern on a $Bi_2Se_3/MoS_2$ 2D heterostructure. The letters "NEU" were "drawn" using different exposure times. (c) Fluorescence microscope image of the same sample (excitation $\lambda$=488 nm). We note that using a focused laser beam, excitons of selected regions could be programmed to recombine radiatively (bright regions) or non-radiatively (dark regions), where the size of the affected area is dependent on the laser spot geometry and recipe used (*i.e.* the power and exposure duration). The smallest "radiative" regions were below a micron in diameter. This method allows for the rapid manipulation and measurement of the interlayer coupling with high spatial resolution, facilitating experiments that produce statistically significant results.

Figure 5 demonstrates a low-cost method to rapidly and precisely manipulate the interlayer coupling of $Bi_2Se_3/MoS_2$ 2D heterostructures with submicron spatial resolution. Figure 5a is a schematic showing how a focused laser beam can be used to apply calibrated energy doses to precisely manipulate the interlayer coupling with high (submicron) spatial resolution. Figure 5b shows the optical image where the method in Figure 5a was used to trace the letters "NEU" using different exposure times, demonstrating a facile method to manipulate the interlayer coupling. The laser-written sample was then imaged using a fluorescence microscope (Figure 5c, $\lambda_{ex}$=488 nm), demonstrating the ability to rapidly measure the degree of interlayer coupling disruption. The ability to rapidly write and read patterns with varying interlayer coupling strength, facilitates experiments that produce statistically significant results. The fluorescence image also demonstrates site-selective light emission with a spatial size close to that of the incident laser spot, enabling photoluminescing pixels (PLPs) tailored down to sub-micron diameters. These site-programmable, color-selectable, atomically-thin, micron-scale PLPs (with effective volumes ~$10^{-21}$ m$^3$) are attractive for optical and optoelectronic applications that require ultra-small form-factors.

Bi$_2$Se$_3$ is well-known for its affinity to oxygen. ARPES measurements have shown that under very low exposures (<0.1 Langmuirs), oxygen inclusion hole-dopes ultraclean Bi$_2$Se$_3$ samples.[55] However, the exposure level in our system is ~$10^9$ orders of magnitude larger, and so we don't expect doping to be a dominant cause behind the observed switching of recombination pathways. Oxygen has been shown to react with Bi$_2$Se$_3$ and form new compounds, such as bismuth oxide;[56], [57] however, the reversible and diffusive behavior at lower exposure ranges suggests the formation of strong chemical bonds, or new materials, is unlikely. Additionally, AFM images of pre and post laser exposure in oxygen do not show changes in the step height or topography, despite the fact the color changed and PL increased dramatically (SI.3b), suggesting the Bi$_2$Se$_3$ was not chemically modified by oxygen. Previous studies measured topographical changes in Bi$_2$Se$_3$ where oxygen modified the structure.[56]

At the same time, oxygen has been shown to easily intercalate between 2D crystals and their substrates, decoupling the two materials and inducing them to behave more "freestanding" (*i.e.* electronically independent).[31], [32], [48]–[52], [62] DFT calculations of a Bi$_2$Se$_3$/MoS$_2$ 2D heterostructure predict that intercalated O$_2$ molecules force the materials apart, enlarging the interlayer separation by 17%, thereby diminishing their interlayer hybridization (SI.6). Furthermore, intercalation is a diffusive process that has been shown to be reversible,[31], [50]–[52] which could explain why laser exposure in an O$_2$-free environment is able to rapidly reverse the changes. The 2D heterostructure appears to have memory of previous laser exposure in oxygen, suggesting more permanent changes are taking place. We found the highly crystalline Bi$_2$Se$_3$ becomes nano- and poly-crystalline when a laser is applied in an oxygen-present environment (SI.1). O$_2$ diffusion might be facilitated through the laser induced break-down of $Bi_2Se_3$ into smaller grains (SI.1), and the subsequent increase in grain boundaries. A schematic explanation of the possible exciton recombination pathways under different conditions is summarized in SI.6.

In conclusion, the interlayer coupling in vertically stacked 2D materials has demonstrated the ability to dramatically alter their properties, and is considered a promising parameter for designing materials with tailored capabilities. The $Bi_2Se_3$/$MoS_2$ 2D heterostructure an ideal platform to study the interlayer interaction of 2D materials because it offers the ability to tune the interlayer coupling *in-situ* and with high-spatial resolution. From a fundamental perspective, in addition to rich excitonic physics, this system interplays strong spin-orbit coupling in non-centro-symmetric crystal structures, and hence could potentially demonstrate novel correlated, spin and valley physics.[63]–[65] In addition, as site-programmable, color-selectable, atomically-thin, micron-scale and intensity-tunable photoluminescing pixels (PLPs), this system could be attractive for ultrathin and flexible optical information storage devices, color converters, micro-cavity-lasers, and other photonic, plasmonic and optoelectronic applications.[66]–[69] The strong oxygen-selectivity of these heterostructures could also be potentially used as low-cost oxygen-sensors and photon/power meters. We also showed data that suggested these materials could be used as $O_2$ storage devices, potentially storing 69 kg/m$^3$ (a factor of 52 times the density of $O_2$ gas at 1 atm).

**Acknowledgement**

Support for this research was received from the National Science Foundation, through grant numbers NSF ECCS 1351424 (a Northeastern University Provost's Tier-1 Seed Grant) and NSF CHE-1764221, as well as the U.S. Department of Veterans Affairs, through the Post-9/11 GI Bill.

**Methods**

Growth of monolayer $MoS_2$ crystals

Monolayer $MoS_2$ was grown using chemical vapor deposition (CVD).[58] The growth setup consisted of quartz tubes [1 inch (2.54 cm) in diameter] in a horizontal tube furnace (Lindberg/Blue M). A quartz boat, containing a thin layer of $MoO_2$ powder (3 mg) with $SiO_2$/Si (MTI Corporation) substrates suspended over the powder with the growth side facing down, was placed in the hot center of the furnace. Sulfur powder (150 mg) was placed near the insulating edge of the furnace upstream. The setup was pumped down and purged with argon gas before it was filled with an Ar atmosphere. Downstream was then opened to atmosphere, in addition to a constant 200 standard cubic centimeter per minute (SCCM) Ar flow. The furnace was heated to different temperatures and at variable rates, depending on the material being grown. The growth was conducted in two stages, 1st-stage and 2nd-stage, where 2nd stage would start once the 1st-stage temperature was reached. See the below table for material specific growth information. After the elapsed time, the furnace was opened and allowed to cool rapidly.

| 2D Crystal | 1st Rate (°C/min.) | 1st Temp. (°C) | 2nd Rate (°C/min.) | 2nd Temp. (°C) | Hold time (min.) |
|---|---|---|---|---|---|
| $MoS_2$ | 50 | 500 | 5 | 712 | 20 |

$Bi_2Se_3$ growth was performed in an identical CVD setup, except a heating wrap was coiled around the quartz tube at the down-stream end, leaving no gap between the furnace and the heating wrap. The $Bi_2Se_3$ powder (50 mg) was placed in the hot center of the furnace. The monolayer $MoS_2$ substrate was placed downstream ~0.75cm from the boundary between the furnace and the heating wrap. The system was pumped down to a base pressure of ~10 mtorr before a 35-SCCM Ar flow was introduced, raising the growth pressure to ~490 mtorr. The heating wrap was set to a temperature of 245°C, and a temperature controller (J-KEM Scientific Model Apollo) ensured it remained within ±2°C. The furnace was heated at a rate of 50°C/min to 530°C and then held there for 20-25 min depending on the desired thickness. Once growth was completed, the furnace was opened and the temperature controller was de-energized, allowing the setup to cool rapidly.

Annealing experiments

All annealing experiments were performed between 240-245°C for 3 hours. All experiments used a flow rate of 3 SCCM, except the annealing under air, where no flow rate was used. The five environments were pure Ar, pure $N_2$, $N_2$+$H_2O^{vapor}$, dry air (21% $O_2$ and 79% $N_2$), and air. All setups, except $N_2$+$H_2O^{vapor}$, were pumped down and filled with the respective gas prior to annealing. The $N_2$+$H_2O^{vapor}$ environment was created by flowing $N_2$ at 3 SCCM, while several boats with deionized $H_2O$ were present in the tube. The heating of the tube caused the $H_2O$ to evaporate. The downstream side "rained" significantly during the entire annealing process and $H_2O$ was still present in most of the boats, verifying that sufficient $H_2O^{vapor}$ was present throughout the annealing process.

Device Fabrication

$Bi_2Se_3$/$MoS_2$ 2D heterostructures grown on 285 nm Si/SiO2 were transferred to an identical chip, that had titanium/gold markers, by PMMA transfer method. First, PMMA C4 was spin coated at 4000 rpm for 60 s and baked 180 C for 1:30 min. Then the chip was immersed in 1 M KOH solution for 4 hours. Obtained PMMA and heterostructure film transferred to new substrate. This was followed by acetone and IPA cleaning to remove PMMA residues.

FET devices were made on 285 nm Si/SiO2 substrate by E-beam lithography using PMMA C4 or A4. The electrodes (5 nm Ti/50 nm Au) were deposited by e-beam evaporator with rate deposition 1 and 3 Å/s, respectively. Lift off process was performed with acetone followed by IPA cleaning.

Instrumentation

Raman and PL spectra were measured using a Renishaw Raman microscope equipped with a 488nm laser and a grating of 1800 lines/mm. A ×100 or ×150 objective focused the laser to diffraction-limited spot size. TEM images and SAED patterns were collected from a JEOL 2010F operated at 200 kV. AFM images were taken from a NanoMagnetics Instruments Ambient AFM. All Raman, PL, AFM, and UV-Vis experiments were performed under ambient condition.

Computational Details

Ab initio calculations were carried out by using the pseudopotential projector augmented-wave (PAW) method [70] implemented in the Vienna ab initio simulation package (vasp) [71], [72] with an energy cutoff of 420 eV for the plane-wave basis set. Exchange-correlation effects were treated using the generalized gradient approximation (GGA) [73], and van der Waals corrections were included using the DFT-D2 method of Grimme [74], where a 7x7x1 Γ-centered k-point mesh was used to sample the Brillouin zone. A large enough vacuum of 15 Å in the z-direction was used to ensure negligible interaction between the periodic images of the films. All the structures were relaxed using a conjugate gradient algorithm with an atomic force tolerance of 0.05 eV/Å and a total energy tolerance of $10^{-4}$ eV. The spin-orbit coupling effects were included in a self-consistent manner.

# Supporting Data

1. TEM Select area electron diffraction (SAED) pattern of a $Bi_2Se_3/MoS_2$ 2D heterostructure

2. Representative device data from a monolayer $MoS_2$ crystal, as well as a $Bi_2Se_3/MoS_2$ 2D heterostructure

3. AFM measurements of pre- and post-anneal and laser exposure in an $O_2$-present environment

4. Results from thermally annealing or focused laser application in different environments, demonstrating $O_2$ is required to modify the interlayer coupling

    a. Table summarizing changes: Energy delivery method *vs.* Environment

    b. Oxygen-induced reversible manipulation of exciton dynamics

    c. Thermally annealed in Ar (*i.e.* an $O_2$-free environment) at 240°C for 3 hours.

    d. Thermally annealed in dry air (*i.e.* 79% $N_2$ + 21% $O_2$)

    e. Thermally annealed in an $N_2+H_2O^{vapor}$ environment

    f. First thermally annealed in air, and then annealed in $N_2$

    g. Thermally annealed in air and then annealed in $N_2$

5. High-control and tunability of heterostructure PL intensity using proper laser-treatment recipe

6. Explanation of possible radiative and non-radiative exciton recombination pathways

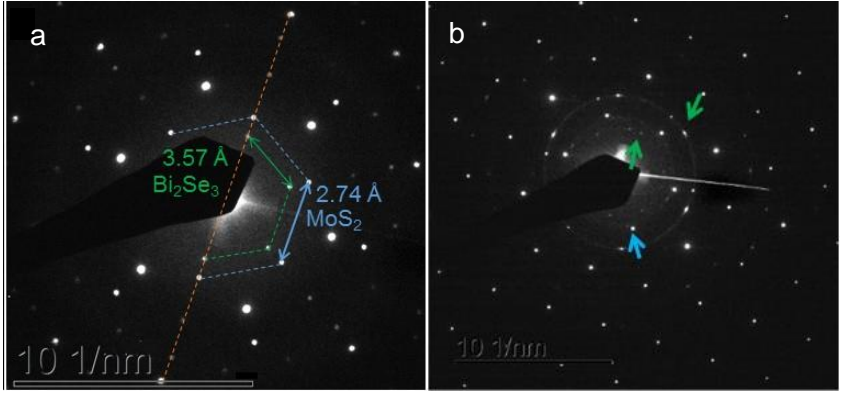

**SI. 1. Transmission electron microscope (TEM) select area electron diffraction (SAED) pattern of a $Bi_2Se_3$/$MoS_2$ 2D heterostructure** (a) $Bi_2Se_3$/$MoS_2$ 2D heterostructure with the lattice parameters labeled of both parent crystals. The distinct dots indicate both materials are highly crystalline. The $Bi_2Se_3$ tends to grow crystallographically aligned (*i.e.* twist angle is 0°). (b) A laser was applied while in an oxygen-present environment (*i.e.* air), and it was found to induce the $Bi_2Se_3$ to become nano- and poly-crystalline. As seen in Figure 3, at higher energy doses the system undergoes more robust changes that are not easily reversed. This is possibly due to the $Bi_2Se_3$ becoming nano-crystalline, which creates more grain boundaries. Oxygen is known to more easily diffuse through grain boundaries (SI.##), which would explain the variation in the PL's rate-of-change.

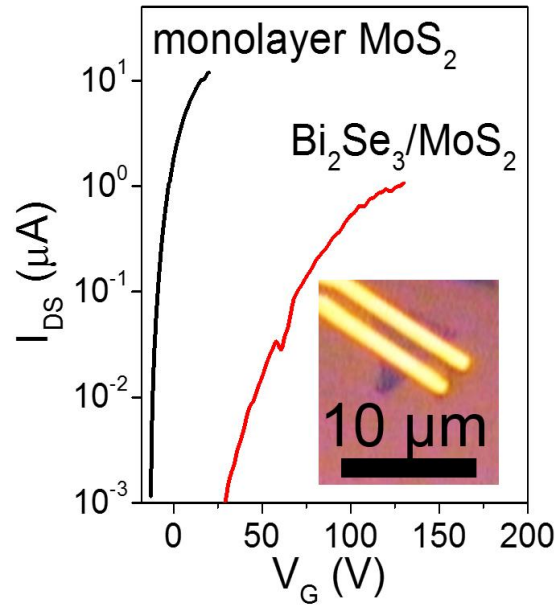

**SI. 2. $Bi_2Se_3$/$MoS_2$ 2D heterostructure device data** Typical drain current ($I_{DS}$) *vs.* gate voltage ($V_G$) in back-gated bare $MoS_2$ and $Bi_2Se_3$/$MoS_2$ devices, respectively. The bare $MoS_2$ device is n-doped into the conduction band, while the right-shifted data-curve of the heterostructure suggests reduction of n-type carriers and lowering of the Fermi level into the band gap.

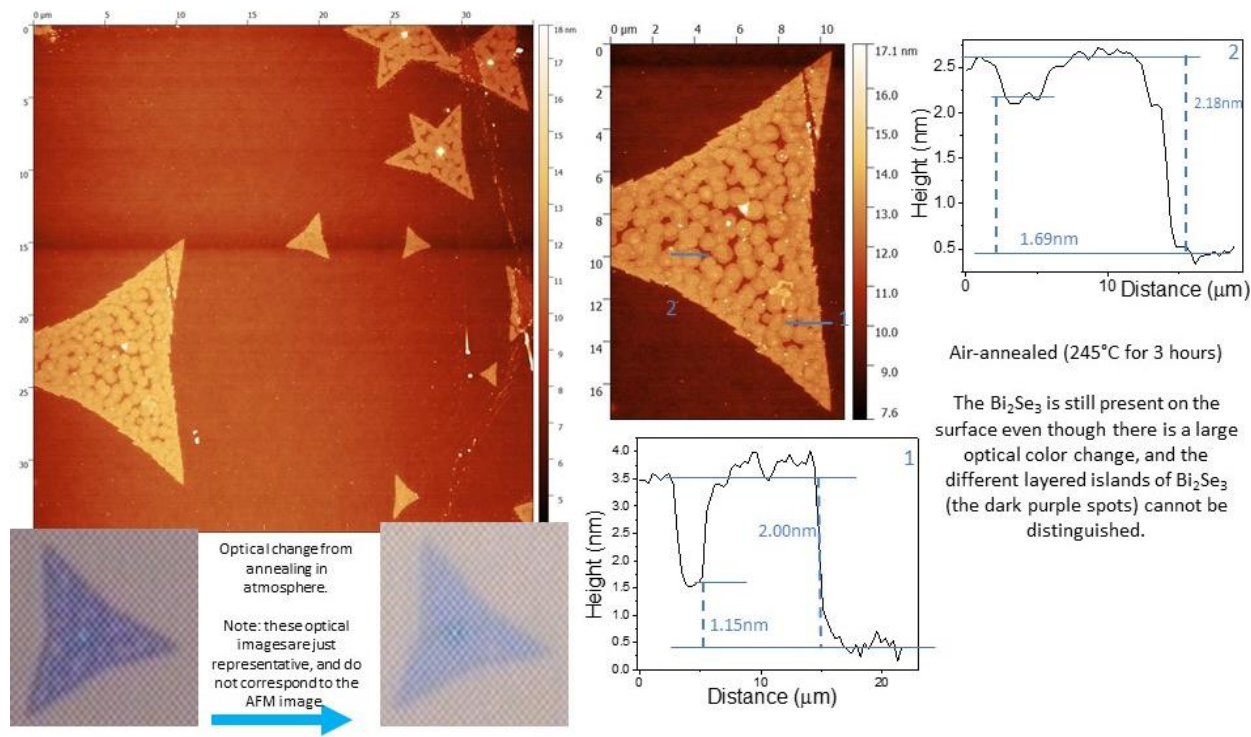

**SI. 3a. AFM images of 2D heterostructure annealed in air.** The images show that annealing in air does not induce obvious changes to the $Bi_2Se_3$ topography or step height despite the fact that there are large optical changes. A $Bi_2Se_3/MoSe_{2-2x}S_{2x}$ 2D heterostructure was used above. Upcoming work will demonstrate that this can be accomplished on a family of monolayer TMDs.

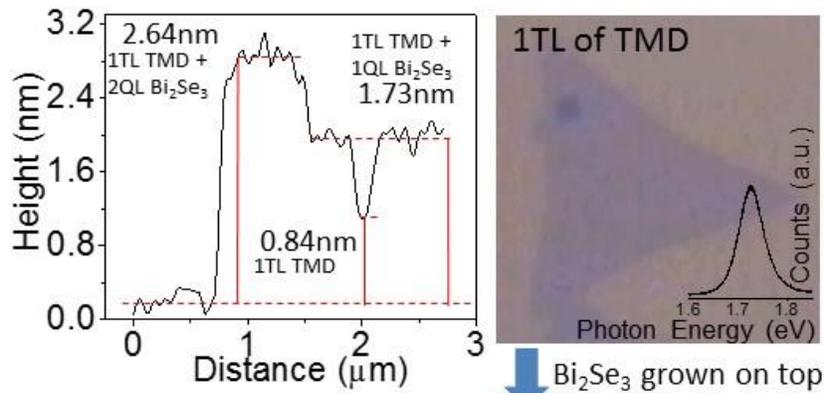

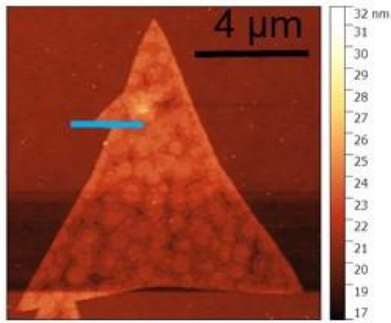
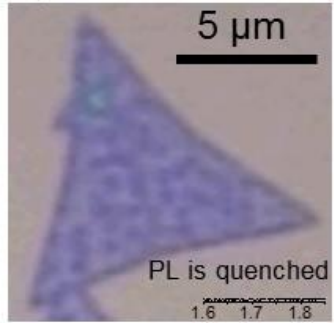

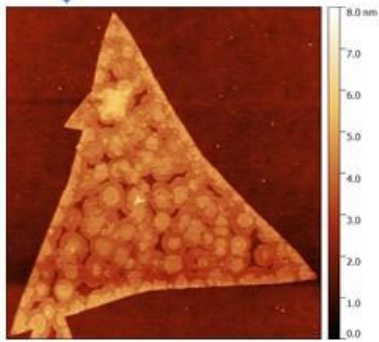
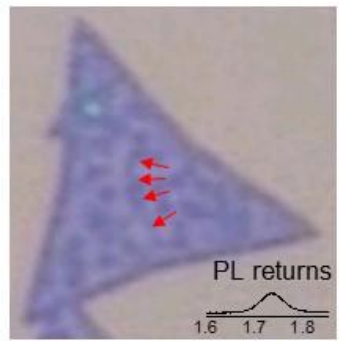

**SI.3b. AFM analysis of 2D heterostructure: as-grown *vs.* air-laser-treated.** Laser exposure in air does not remove the $Bi_2Se_3$, a key finding demonstrating the exciton recombination pathway switching is more subtle and not simply because the $Bi_2Se_3$ is being removed. Interestingly, the effect of the laser exposure on the topography is difficult to discern, where the pre- and post-air-laser-treatment AFM images look nearly identical. No clear changes were detected, despite the fact that the optical and photoluminescence properties are drastically altered. A $Bi_2Se_3/MoSe_{2-2x}S_{2x}$ 2D heterostructure was used above. Upcoming work will demonstrate that this can be accomplished on a family of monolayer TMDs.

| Surrounding environment vs. Energy delivery method: How the PL photon count, optical contrast, and perceived color are affected | | Energy delivery method | | | | | |
|---|---|---|---|---|---|---|---|
| | | Focused laser beam (488nm) | | | Thermal Annealing (243°C) | | |
| | | PL | Contrast | Color | PL | Contrast | Color |
| Surrounding environment | Air ($N_2+O_2+H_2O$) | ↑ | ↓ | Brightens | ↑ | ↓ | Brightens |
| | Dry air (79% $N_2$ + 21% $O_2$) | | | | ↑ | ↓ | Brightens |
| | Wet $N_2$ ($N_2+H_2O$) | | | | ↓ | ↑ | Darkens |
| | $N_2$ | ↓ | ↑ | Darkens | ↓ | ↑ | Darkens |
| | Ar | | | | ↓ | ↑ | Darkens |

**SI.4a. Table summarizing the changes induced depending on the energy delivery method and the environment.** There are two energy delivery methods (Focused laser beam and Thermal annealing), and five different environments (Air, Dry air, Wet $N_2$, $N_2$, and Ar). The results show that $O_2$ is required to switch the exciton recombination pathway from non-radiative to radiative.

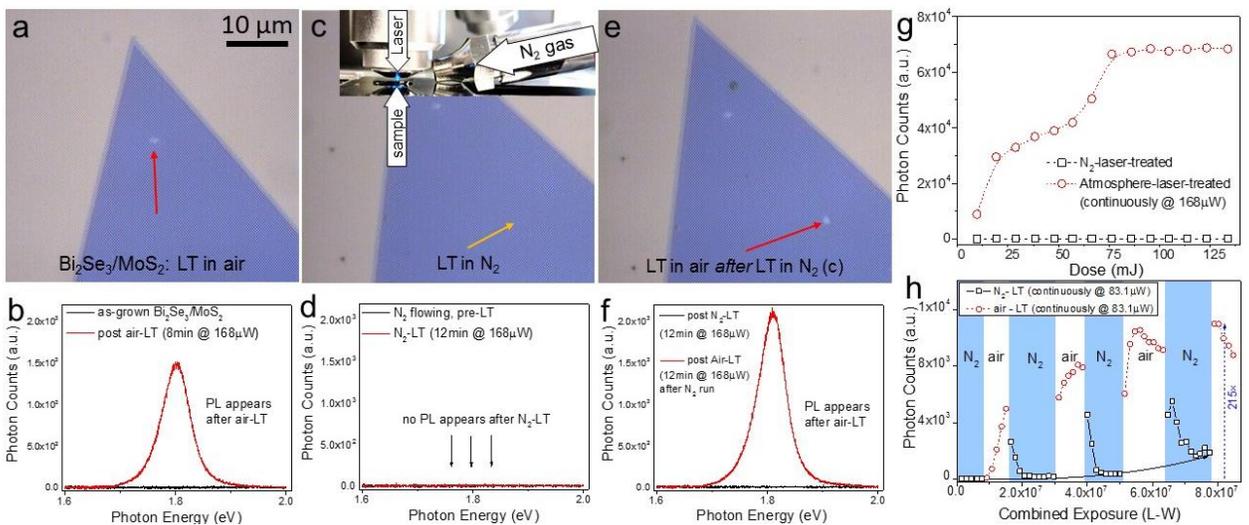

**SI. 4b. Oxygen-induced reversible manipulation of exciton dynamics** (a) Optical image of a $Bi_2Se_3/MoS_2$ heterostructure. The arrow indicates a spot whose color changed after being air-laser-treated (*i.e.* LT in air). (b) PL spectra from the same spot before and after the air-laser-treatment, demonstrating the predicted behavior (*i.e.* PL recovery). (c) A different location (orange arrow) on the same sample was $N_2$-laser-treated by flooding the environment with $N_2$ gas flow and displacing the air. $N_2$-laser-treatment inhibits the color change observed in (a). Inset shows the setup used. (d) PL spectra from before and after $N_2$-laser-treatment, showing no perceivable appearance of PL after $N_2$-laser-treatment. Laser-treatment and annealing experiments were separately performed in synthetic dry $O_2+N_2$ mixture and moisture-saturate $N_2$ (SI.1) from which any role of the other components of air, *e.g.* $N_2$, $H_2O$, $CO_2$ could be eliminated. (e) The same spot in (c) was air-laser-treated. The color-change is perceivable, along with the appearance of the PL spectrum, shown in (f), demonstrating that extended $N_2$-laser-treatment had no obvious deleterious effect on the switching of the exciton recombination pathways. (g) Comparison of air-laser-treatment *vs.* $N_2$-laser-treatment using a high-power recipe. The PL intensity grew over an order of magnitude under air-laser-treatment, compared to the flat-growth under $N_2$-laser-treatment. (h) Variation of PL intensity under alternating air- and $N_2$-laser-treatments, showing that $N_2$-laser-treatment diminishes the PL intensity several decades. The solid black arrow shows how the baseline PL reading monotonically grows after repeated cycling of the laser-treatments, and the PL's rate-of-change between air-laser-treatments varies, suggesting $N_2$-laser-treatment may not return the heterostructure to its as-grown state. The dashed blue arrow shows the overall PL intensity growth factor, up to 215×, achievable by this cycling approach. Taken together, this demonstrates an unprecedented degree of controlled manipulation of PL achievable in a site-selectable manner.

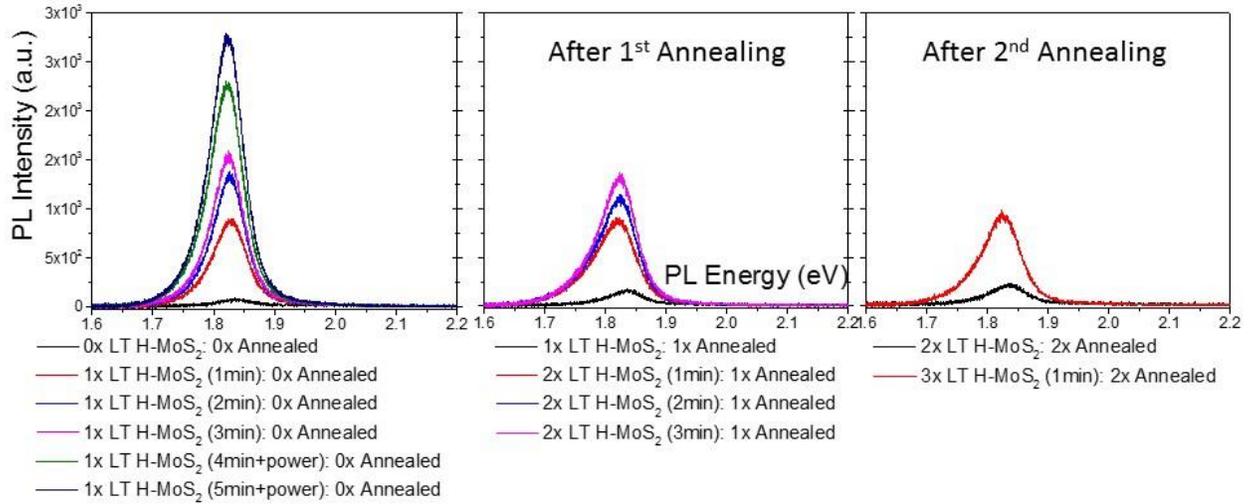
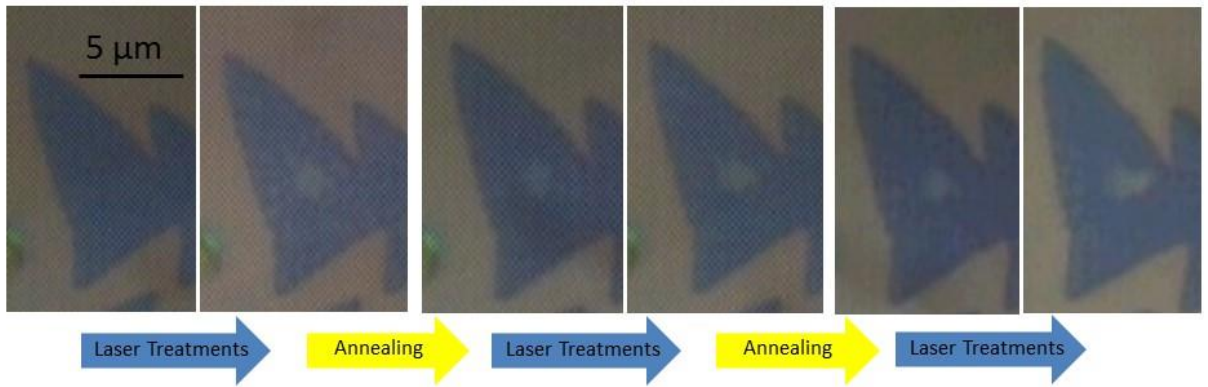

**SI. 4c. 2D heterostructures annealed in Ar (*i.e.* an O$_2$-free environment) at 240°C for 3 hours.** 1-3 layers of Bi$_2$Se$_3$ were grown on monolayer MoS$_2$ using vapor-phase deposition. The photoluminescence was not only quenched after each annealing session, but it was also controllably increased using laser-treatment in air.

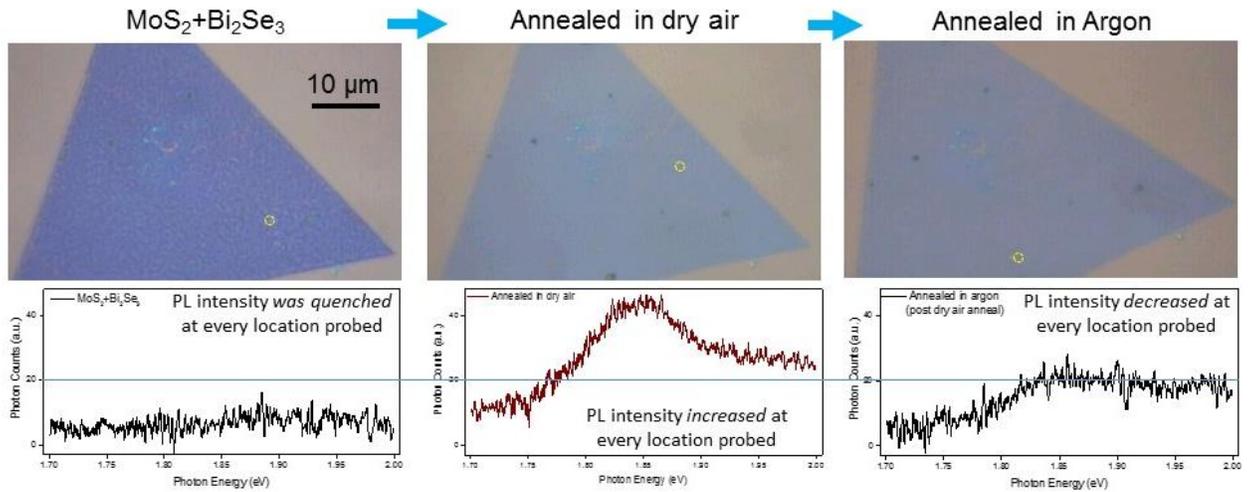

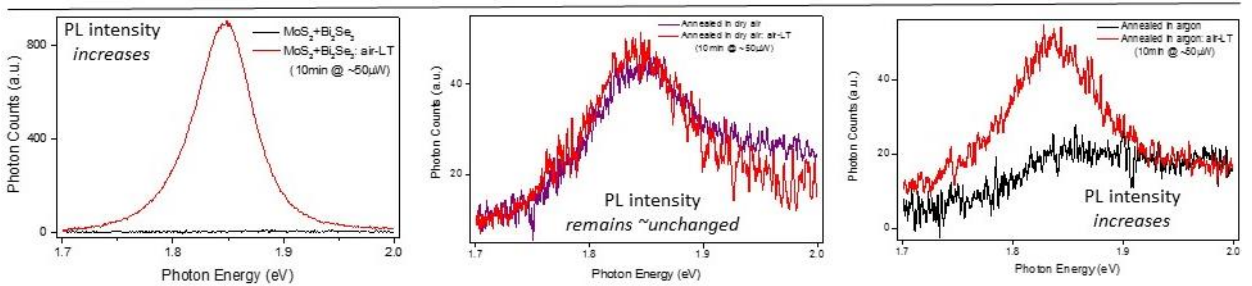

**SI. 4d. 2D heterostructure annealed in dry air (*i.e.* 79% $N_2$ + 21% $O_2$).** The behavior is similar to that of the air-annealing experiments in that the PL intensity increases when dry-air-annealed and then decreases again after $N_2$-annealing. However, it is different in two subtle points: (1) after dry-air-annealing, air-laser-treatment does not affect the PL; and (2) a possible new PL signal emerges in the upper energies (~2.0eV), where the PL plateaus to a higher intensity than that seen at 1.7eV (*i.e.* the PL spectra is not symmetric, showing higher intensities at higher energies).

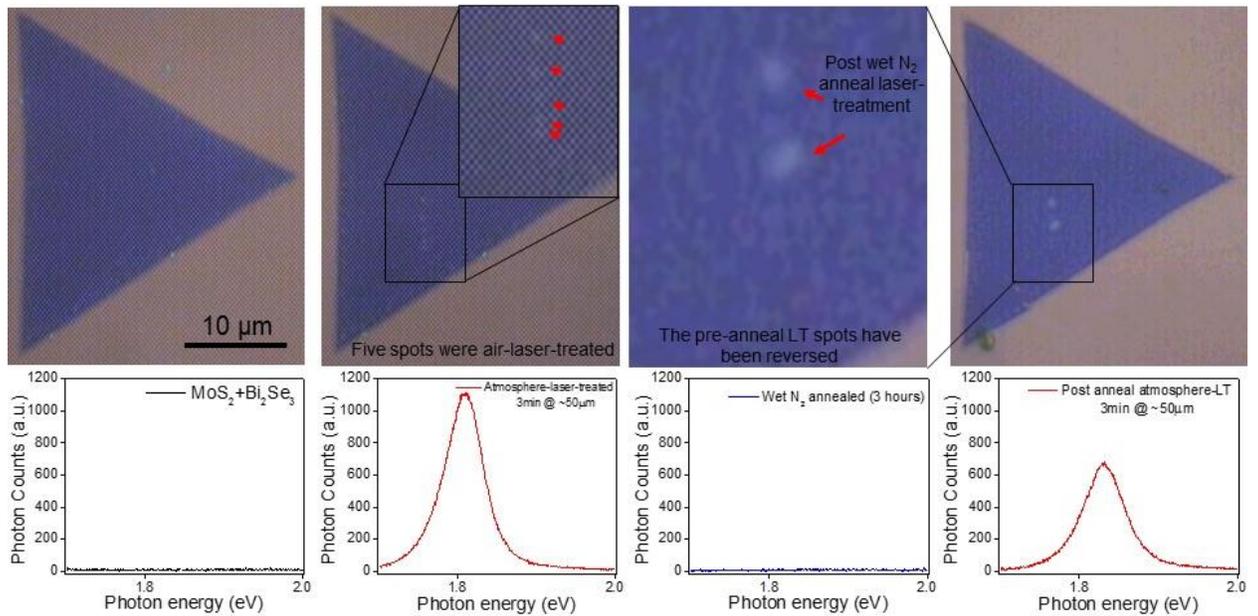

**SI. 4e. 2D heterostructure annealed in an $N_2$+$H_2O^{vapor}$ environment.** This experiment established that water vapor does not react with the heterostructures to induce the optical and photoluminescence changes observed during air-laser-treatment in air (*i.e.* water vapor is not necessary to switch exciton recombination pathway). In fact, the heterostructure will reverse both the air-laser-treatment induced optical and photoluminescence changes, when annealed in an $N_2$ environment saturated with water vapor. As seen above, five locations were air-laser-treated prior to annealing, and an optical change was induced in each on. Annealing reversed the optical change in all five spots. The two larger spots seen in the right two optical images were induced post-anneal, demonstrating it was still possible to switch the recombination pathway. A $Bi_2Se_3$/$MoSe_{2-2x}S_{2x}$ 2D heterostructure was used above. Upcoming work will demonstrate that this can be accomplished on a family of monolayer TMDs.

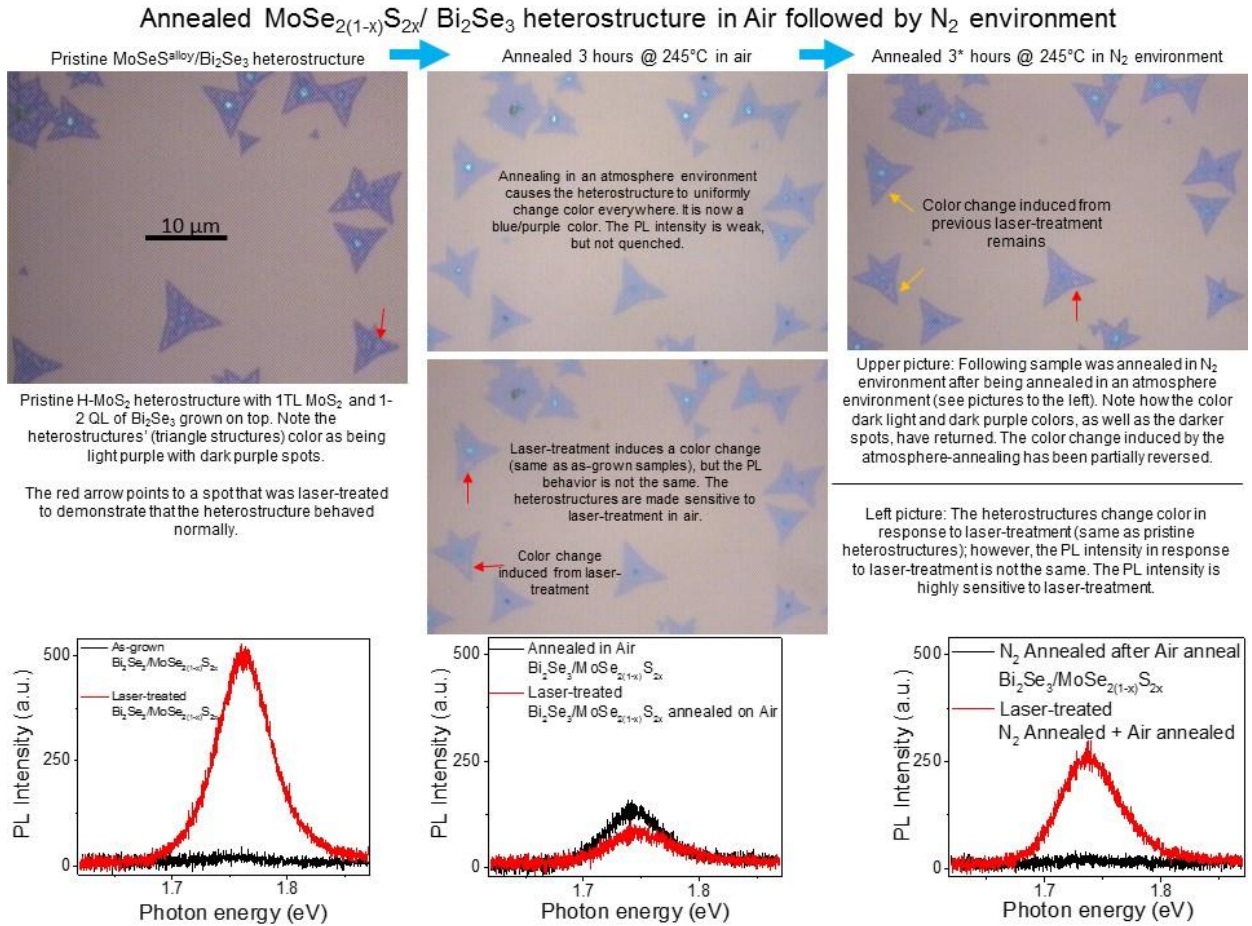

**SI. 4f. 2D heterostructures first annealed in air, and then annealed in N$_2$. <u>PL spectra at Different locations.</u>** The heterostructures are monolayer MoSe$_{2-2x}$S$_{2x}$ with 1-2 layers of Bi$_2$Se$_3$ CVD grown on top. (**Top Optical Images**) Note how the air-annealed heterostructures appear a lighter shade of purple and have less contrast, compared to the as-grown heterostructures. The heterostructures were then N$_2$-annealed, resulting in the air-annealed optical changes being partially reversed (*i.e.* there is a darker shade of purple and more contrast). (**Bottom PL spectra**) Left to right are the as-grown, air-annealed, and N$_2$-annealed (following air-annealed), respectively. Each graph has the pre- and post-air-laser-treated PL spectra. The data between annealing sessions was taken at different locations (specified by the red arrows). Air-annealed samples will have a brighter PL than the as-grown samples, but it will be weaker than a laser-treated as-grown heterostructure. The PL intensity of air-annealed heterostructures will *decrease* when air-laser-treated. A Bi$_2$Se$_3$/MoSe$_{2-2x}$S$_{2x}$ 2D heterostructure was used above. Upcoming work will demonstrate that this can be accomplished on a family of monolayer TMDs. A Bi$_2$Se$_3$/MoSe$_{2-2x}$S$_{2x}$ 2D heterostructure was used above. Upcoming work will demonstrate that this can be accomplished on a family of monolayer TMDs.

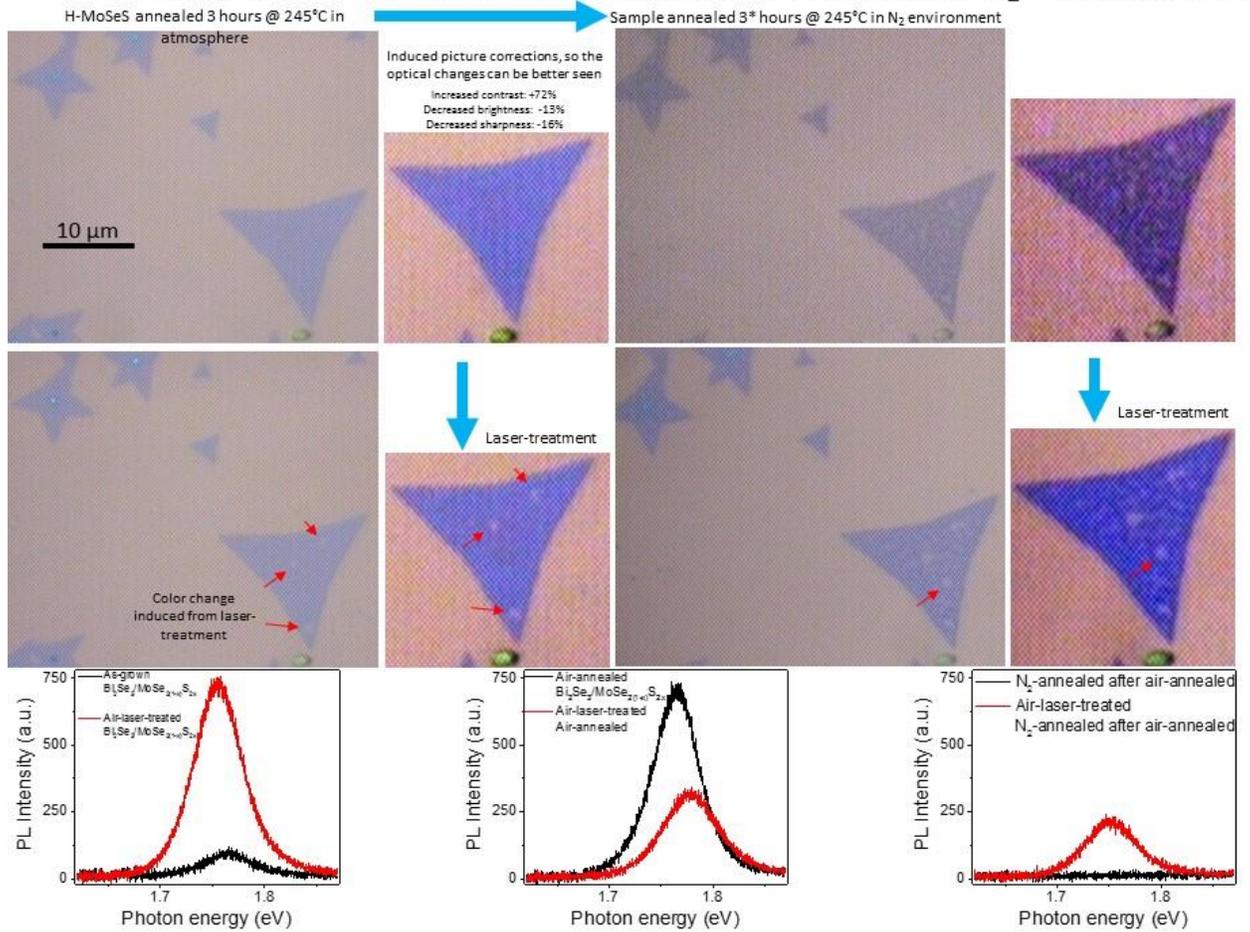

**SI. 4g. 2D heterostructures annealed in air and then annealed in N₂. PL spectra at the same location.** The heterostructures are monolayer MoSeS with 1-2 layers of Bi₂Se₃ CVD grown on top. (**Bottom PL spectra**) The data between annealing sessions was taken at the same location. The air-laser-treated location maintained approximately the same intensity; however, the peak blue-shifted. The PL intensity of air-annealed heterostructures will *decrease* when air-laser-treated, which is the same behavior as in SI.1c.i. After N₂-annealing, the PL will be quenched, and air-laser-treatment induces the PL upward, which is the same behavior observed in as-grown heterostructures. A Bi₂Se₃/MoSe$_{2-2x}$S$_{2x}$ 2D heterostructure was used above. Upcoming work will demonstrate that this can be accomplished on a family of monolayer TMDs.

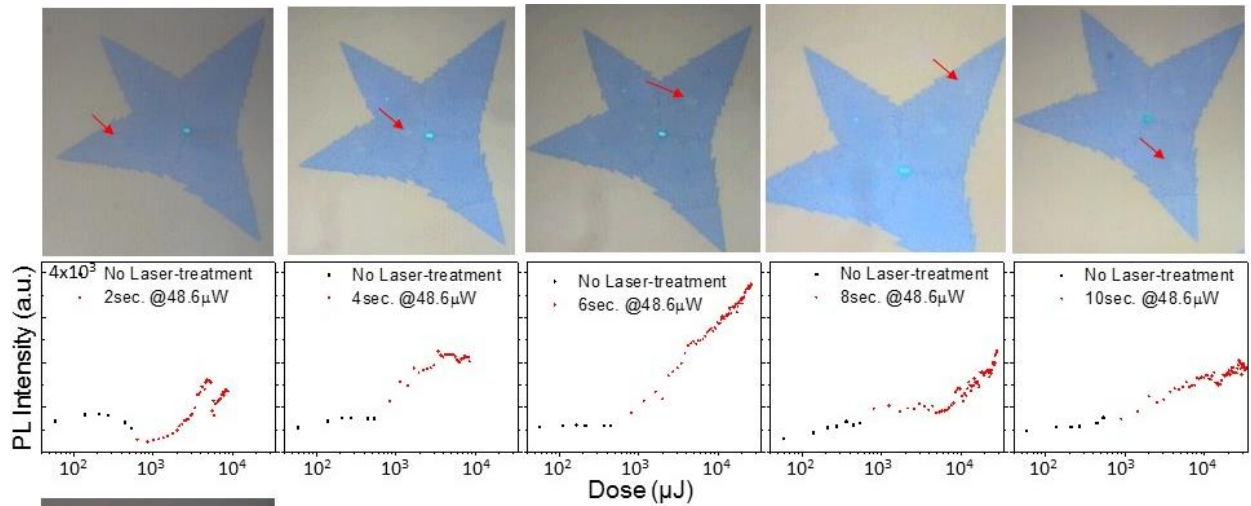

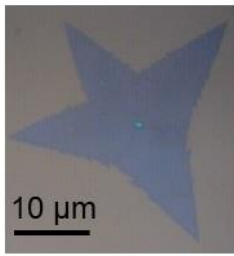

As-grown Bi$_2$Se$_3$/MoS$_2$ heterostructure

**SI. 5. High Tunability and control of interlayer coupling and PL intensity. Rate of change of PL is dependent on the energy application recipe.** A Bi$_2$Se$_3$/MoS$_2$ crystal was air-laser-treated using five different recipes, where the power was held constant power, but time was changed. Different locations on the same Bi$_2$Se$_3$/MoS$_2$ crystal were studied, allowing the results to be better compared. The results show that 6 sec. at 48.6μW produced the most consistent results.

The difference in curve shapes is due to competing factors: too much air-laser-treatment results in degradation of the heterostructure; however, too little power does not induce sufficient change to be permanent. Additionally, using very low powers (~0.91μW) causes the PL intensity to decline. We believe this is because the powers perturb the heterostructure into a better coupling, which squeezes the intercalated oxygen out. It has been shown previously that changing the interlayer coupling will change the rate of diffusion of oxygen intercalating between graphene and the surface (see [2] and SI.9).

Section 6 – Explanation of possible radiative and non-radiative exciton recombination pathways

This family of 2D heterostructures has produced intriguing data that speaks both to the promising applications, as well as the complexity of the underlying science. There are a number of established mechanisms in peer-reviewed literature that may apply because they are in agreement with portions of the data. To the best of our knowledge, our system has multiple competing mechanisms that are taking place to produce the observed behavior.

Below we describe the strengths and weaknesses of a variety of mechanisms that could be present, and then highlight the ones we believe play a primary role. To best convey our thought process, we list all the key observations (KOs), followed by the list of the mechanisms we believe could be present.

Key Observations

1. The PL of the TMD is over 99% quenched when only *one layer* of $Bi_2Se_3$ is CVD grown on the TMD. One layer of $Bi_2Se_3$ is not thick enough to reflect or absorb the incoming and outgoing photons, meaning the $Bi_2se_3$ introduces a non-radiative electron-hole recombination path.
2. Several changes are induced from air-laser-treatment:
    a. PL intensity will increase.
    b. PL intensity can be precisely tuned (*i.e.* recovered with high control) over several orders of magnitude.
    c. Affected regions undergo a perceived color change, appearing brighter and with less contrast, compared to as-grown heterostructures.
    d. PL peak position is in the same location as the monolayer TMD's PL peak position from pre-$Bi_2Se_3$ growth, strongly indicating that the recovered PL observed is from the excitons and trions in the TMD. This would indicate that air-laser-treatment removes the non-radiative electron-hole recombination pathway, allowing the excitons and trions to recombine at the K-point in the TMD, where it is direct bandgap.
    e. Laser-treatment recipe (*i.e.* laser-power and time interval) will affect the rate of change of the PL intensity and peak position shift.
3. Air-laser-treatment changes can be spatial controlled with sub micrometer precision (*i.e.* the laser spot size).

4. $O_2$ is required to be present in the surrounding environment for changes to be induced, meaning it reacts with the heterostructures.
5. Air-laser-treatment induced changes can be reversed by thermally annealing or laser-treating the heterostructure in an $O_2$-free environment.
6. $Bi_2Se_3$/$MoS_2$ heterostructures are p-type, and p-doped compared to pristine monolayer $MoS_2$.
7. Air-laser-treating or annealing does not remove the $Bi_2Se_3$ from the surface, as seen by AFM scans.
8. Larger air-laser-treatment doses appear to induce permanent changes that make the heterostructure more sensitive in subsequent air-laser-treatments.
9. $Bi_2Se_3$ grows crystalline and with long-range order on the TMD, suggesting strong van der Waals epitaxy-mediated growth between the two component layers, and that they couple together.
10. Density functional theory (DFT) calculations of the $Bi_2Se_3$/$MoS_2$ 2D heterostructure predict that intercalated $O_2$ molecules will increase the interlayer separation, disrupt the interlayer bonding, and diminish the interlayer interaction, thereby inducing the two materials to behave more "free-standing".

Exciton recombination pathways that may be present

#1 –A straddled or staggered bandgap induces the photoluminescence quenching

Well-coupled 2D heterostructures have been shown to have overlapping bandgaps,[75], [76] which we believe is happening when $Bi_2Se_3$ is grown on the TMDs, for reasons explained in key observation 9 (KO-9). There are three different types of overlapping bandgaps: broken, straddled, and staggered. Device data showed that $Bi_2Se_3$/$MoS_2$ is p-type, meaning the bandgap is not broken (broken bandgaps behave metallic), but is forming either a straddled or a staggered bandgap. All four TMD's studied in this work are wide bandgap compared to $Bi_2Se_3$ (>1.6eV to ~0.3eV),[55] meaning both a straddled and a staggered bandgap would introduce a non-radiative electron-hole recombination pathway.

Mechanism #1 is in agreement with the PL quenching (KO-1) and why Bi2Se3/MoS2 is p-type (KO-6).

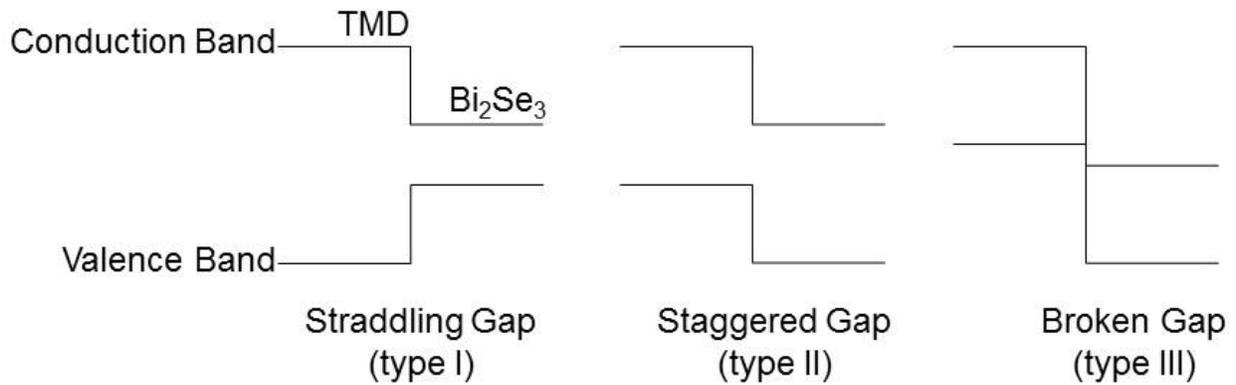

**SI. 6a. Possible non-radiative electron-hole recombination pathways to explain the PL quenching observed in as-grown heterostructures.** Both straddled and staggered bandgaps are in agreement with the data and explain the photoluminescence quenching observed because they introduce a non-radiative recombination pathway. Although a broken bandgap also has a non-radiative recombination pathway, it disagrees with KO-6; broken band gaps are metallic, whereas our device was p-type.

#2 – $O_2$ intercalates between TMD and $Bi_2Se_3$ and diminishes the interaction strength and coupling

As the heterostructure is air-laser-treated, it undergoes numerous dramatic optical and PL changes; however, the $Bi_2Se_3$ is not removed (KO-7), as shown with AFM. We know that whatever the mechanism is, it is highly local (KO-3) and that $O_2$ is required for the process (KO-4).

We believe that as the heterostructure is laser-treated or annealed in an $O_2$-present environment, $O_2$ intercalates into the interlayer spacing of the heterostructure (*i.e.* between the $Bi_2Se_3$ and TMD layers), where it disrupts the interlayer coupling, leading to the changes in properties observed. It has been shown that $O_2$ intercalating between 2D materials and their surface decouples the 2D material, making quasi free-standing.[31], [32], [49]–[52], [77] In this case, it decouples the materials, allowing the TMD to regain the radiative recombination pathway.

Additionally, each $O_2$ molecule is only able to disrupt the interlayer coupling locally, meaning the amount of disruption in a certain area is dependent on the number of $O_2$ molecules; the more $O_2$ there is, the greater the interlayer coupling disruption. This mechanism (or radiative recombination pathway) is in agreement with numerous key observations, stated below:

KO-2a & KO-2d: As the interlayer coupling is disrupted, the non-radiative e-h recombination path will no longer be allowed, permitting e-h excitonic pairs to radiatively recombine in the TMD.

KO-2b: The amount of recovery is dependent on the amount of $O_2$ that intercalates, explaining the high control.

KO-f: Changing the air-laser-treatment recipe will affect the diffusion of $O_2$, thereby affecting the rate of change.

KO-e: As $O_2$ diffuses into the interlayer region, the environment surrounding the TMD will change, thereby altering the surrounding dielectric constant. All exciton quasiparticles emit electric field lines that affect the quasiparticle's properties, and in 2D materials, these lines exist outside of the material, making the excitons and trions highly sensitive to the surrounding environment. By increasing the dielectric constant, one increases the binding energy, thereby lowering the peak position.[46]

KO-c: The perceived color change indicates the material is becoming more transparent, which could be because the interlayer bandgap is removed as the materials decouple.

KO-5: The changes can be reversed by annealing in an $O_2$ free environment because the $O_2$ molecules diffuse out.

KO-8: It has been shown that as $Bi_2Se_3$ is air-laser-treated, the $Bi_2Se_3$ will break-up into small grains. It is likely that increasing the number of grain boundaries increases the rate of $O_2$ diffusion, a fact that has been shown previously for graphene and silicene on various metal substrates.[31], [32], [49], [50], [52], [77]

KO-10: DFT calculations predict intercalated $O_2$ will disrupt the interlayer interaction.

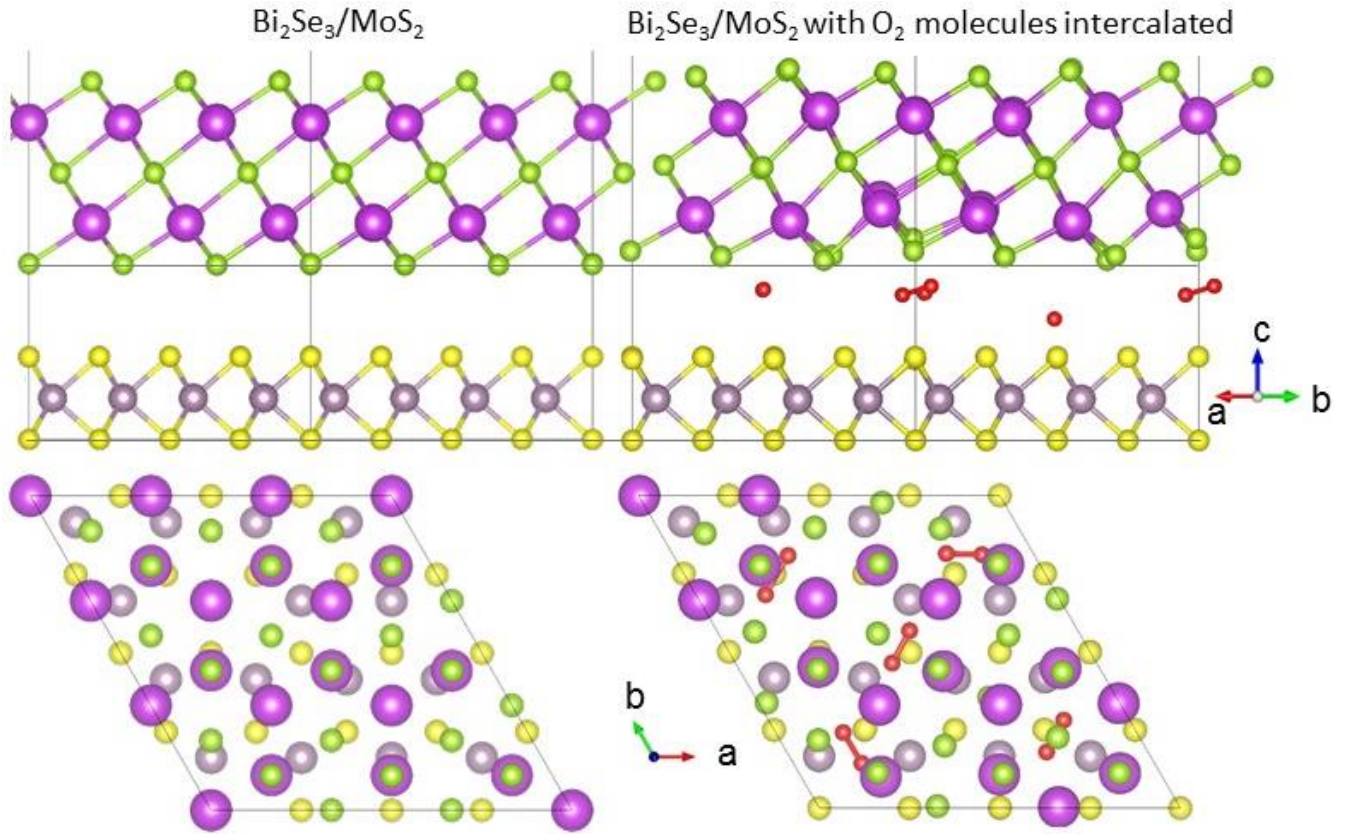

**SI. 6b. DFT calculations comparing a Bi$_2$Se/MoS$_2$ superlattice with and without O$_2$ intercalated.** The calculations predict that O$_2$ intercalation will increase the interlayer separation, disrupt the interlayer bonding, and diminish the interlayer interaction. Note how the interlayer separation increases after adding O$_2$ molecules, going from an average separation of 3.57Å to an average separation of 4.18Å, a 17% increase. Interestingly, the O$_2$ molecules create an uneven landscape of selenium atoms at the interface, pushing different atoms to different separation values. The above 2D heterostructure is rotationally aligned (*i.e.* twist angle is 0°), where 3 Bi$_2$Se$_3$ unit cells are the same length as 4 MoS$_2$ unit cells.

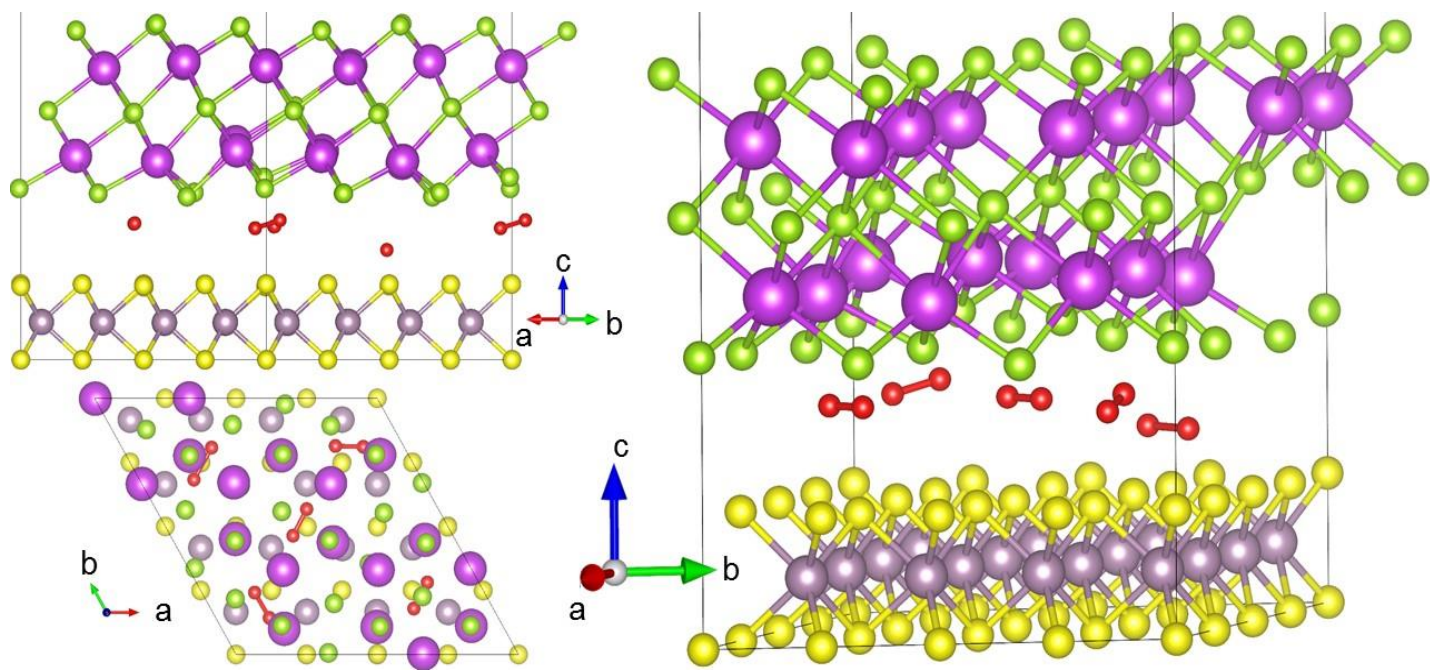

**SI. 6c. DFT calculations showing the location of the O$_2$.** These atomically-thin layers may be excellent candidates for oxygen storage devices, potentially storing 69 kg/m3 ¬(a factor of 52 times the density of O2 gas at 1 atm). The Bi$_2$Se$_3$/MoS$_2$ superlattice has a predicted volume of 1.936nm$^3$ (1.247x1.086x1.43 nm), and the combined mass of five O$_2$ molecules is 1.328x10$^{-25}$ kg, yielding a density of 68.6 kg/m$^3$.

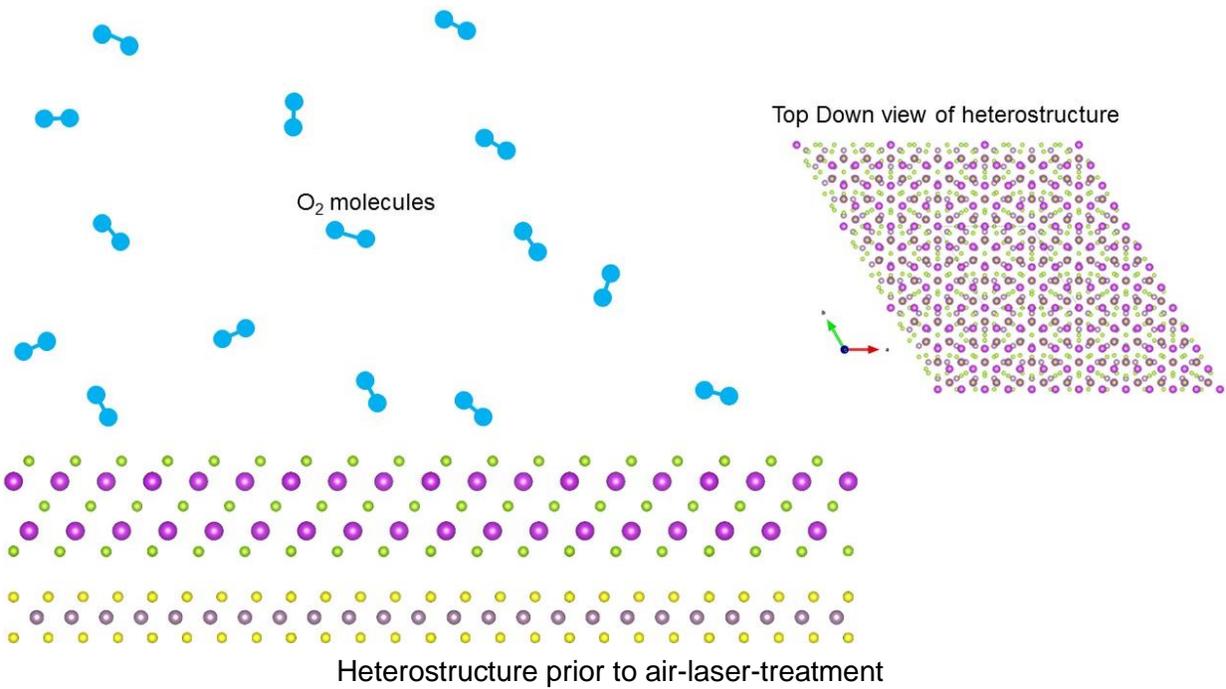

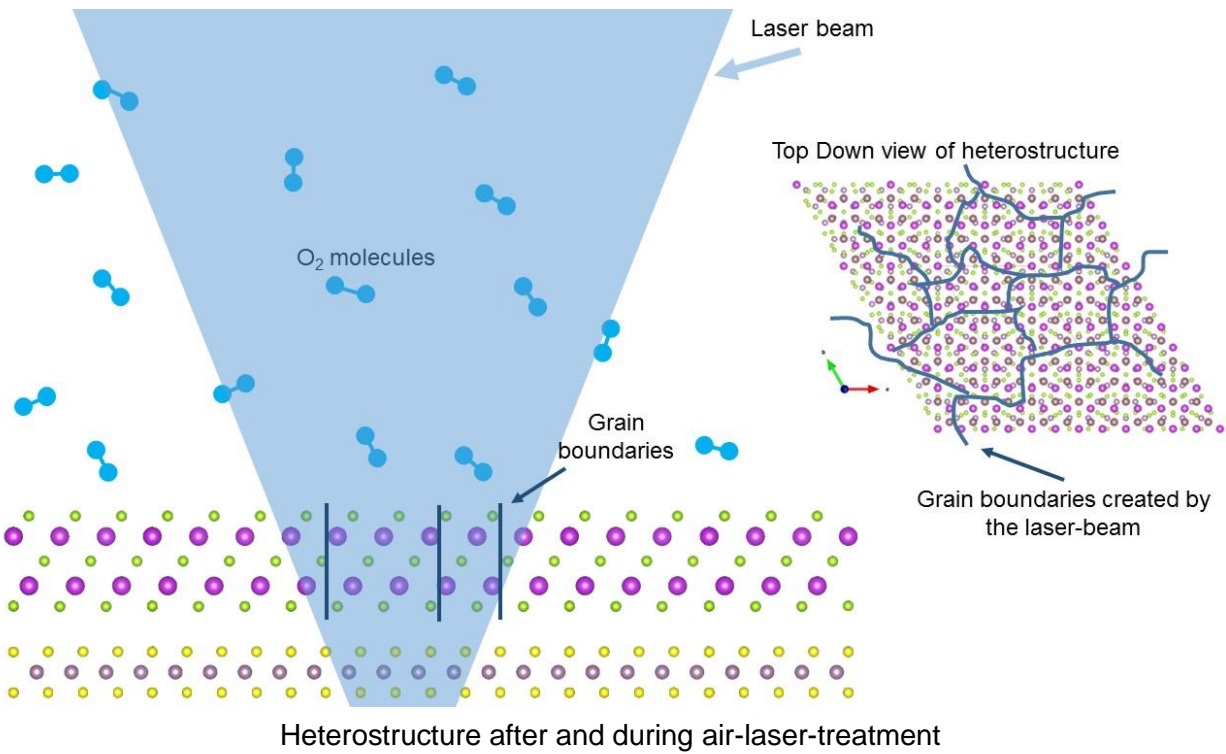

**SI. 6d. Diagrams demonstrating the creation of grain boundaries, facilitating O₂ intercalation into the interlayer region.** The upper diagram shows the heterostructure in an O₂ environment prior to laser-treatment, and the lower diagram is during laser-treatment.